\documentstyle[12pt]{article}

\input{psbox.tex}

\def\be{\begin{equation}}
\def\ee{\end{equation}}
\def\bea{\begin{eqnarray}}
\def\eea{\end{eqnarray}}

\def\br{\begin{thebibliography}{abc}}
\def\er{\end{thebibliography}}

\def\a{\alpha}
\def\b{\beta}

\def\d{\delta}

\def\l{\lambda}
\def\m{\mu}
\def\n{\nu}
\def\o{\omega}
\def\p{\pi}

\def\L{\Lambda}
\def\O{\Omega}
\def\P{\Pi}

\def\pa{\partial}

\def\bar#1{\overline{#1}}

\def\Hat#1{\rlap{\kern.10em$\widehat{\phantom G}$}#1}
\def\HAt#1{\rlap{\kern.05em$\widehat{\phantom G}$}#1}

\def\cAp#1{\rlap{\kern.1em$\widehat{\phantom{G\vrule height.8em}}$}#1{}}
\def\Cap#1{\rlap{\kern.05em$\widehat{\phantom{G\vrule height.8em}}$}#1{}}

\newcommand{\sect}[1]{\setcounter{equation}{0}\section{#1}}
\newcommand{\subsect}[1]{\subsection{#1}}

\def\sxn#1{\bigskip\bigskip \sect{#1} \medskip}
\def\subsxn#1{\bigskip \subsect{#1} \medskip}

\begin{document}

\thispagestyle{empty}
\setcounter{page}{0}

\begin{flushright}
SU-4240-574 \\
November 1994
\end{flushright}
\vspace*{15mm}
\centerline {\LARGE EDGE STATES IN GAUGE THEORIES: }
\vspace{5mm}
\centerline {\LARGE THEORY, INTERPRETATIONS AND PREDICTIONS}
\vspace*{15mm}
\centerline {\large A. P. Balachandran$^{1}$, L. Chandar$^{1}$, E.
Ercolessi$^{2}$}

\vspace*{5mm}
\centerline {\it $^{1}$ Department of Physics, Syracuse University,}
\centerline {\it Syracuse, NY 13244-1130, U.S.A.}
\centerline {\it $^{2}$  Dipartimento di Fisica, Universit$\bar{\mbox{a}}$
di Bologna,}
\centerline {\it Via Irnerio 46, 40138 Bologna, Italy.}

\vspace*{25mm}
\normalsize
\centerline {\bf Abstract}
\vspace*{5mm}
Gauge theories on manifolds with spatial boundaries are studied.  It
is shown that observables localized at the boundaries (edge observables) can
occur in
such models irrespective of the dimensionality of spacetime.  The intimate
connection of these observables to charge fractionation, vertex operators and
topological field theories is described.  The edge
observables, however, may or may not exist as well-defined operators in a fully
quantized theory depending on the boundary conditions
imposed on the fields and their momenta.  The latter are obtained by requiring
the Hamiltonian of the theory to be self-adjoint and positive definite.  We
show that these
boundary conditions can also have nice physical interpretations
in terms of certain experimental parameters such as the penetration depth of
the
electromagnetic field in a surrounding superconducting medium.  The
dependence of the spectrum on one such parameter is explicitly exhibited for
the Higgs model on a spatial disc in its London limit.  It should be possible
to test
such dependences experimentally, the above Higgs model for example being a
model for a superconductor.  Boundary conditions for the 3+1 dimensional $BF$
system confined to a spatial ball are studied.  Their physical meaning is
clarified and their influence on the edge states of this system (known to exist
under certain conditions) is discussed.  It is pointed out that edge states
occur for topological solitons of gauge theories such as the 't Hooft-Polyakov
monopoles.

\newpage

\baselineskip=24pt
\setcounter{page}{1}

\sxn{Introduction}

The study of gauge fields on spatial manifolds with boundaries has
attracted attention in recent times. A strong incentive for these
investigations comes from attempts to model Quantum Hall Effect
(QHE) where edge states localized at spatial boundaries are known to
exist and to be of fundamental physical significance \cite{qhe}. A second
reason for these investigations is the remarkable relation of these
states to conformal field therories (CFT's) and the possibility
thereby suggested of a three-dimensional approach to these
two-dimensional theories \cite{witt,csb}.

In previous work, we have examined Chern-Simons (CS) \cite{csb},
Maxwell-Chern-Simons (MCS) \cite{mcs} and Higgs field dynamics \cite{higgs}
with disk as
spatial manifold and reproduced many previously known results on
edge states. It was also established that much of the previous work is
limited in scope and fails to exhaust all available physical
possibilities. This is because quantization depends on certain
boundary conditions (BC's). The latter, for a Hamiltonian bounded
below, and with suitable additional assumptions, can be parametrized by
a non-negative parameter $\lambda$ for the CS and MCS cases, and by
two nonnegative parameters $\lambda$ and $\mu$ for the Higgs case.
All work we know of prior to \cite{mcs} had tacitly assumed that
$\lambda=0$. In \cite{mcs}, we had also announced an interpretation of
$\lambda$: if the disk is surrounded by a superconductor, then
$1/\lambda$ is proportional to the penetration depth into its
ambient medium. That is, $\lambda$ is proportional to the vector meson mass in
the surrounding medium.

In this paper, we will explain edge states using a few elementary
examples and then will establish the interpretation of $\lambda$
stated above . [We do not however have a good interpretation of $\mu$.] With
this meaning at hand, it is easy to imagine experiments to
observe the effects of  $\lambda$, as we will see. Further,
we will extend our previous work on edge states of the $BF$ system in three
spatial dimensions \cite{pao} by enlarging the boundary conditions
contemplated.
They too are characterized by parameters like $\lambda$ and $\mu$
with properties similar to those in two dimensions.

Section 2 initiates our considerations by introducing edge states
using Maxwell's theory in $1+1$ spacetime, the spatial slice being
the interval $[-\frac{L}{2},\frac{L}{2}]$. Although the quantum
physics of this model is simple, it is nevertheless rich enough to
display several phenomena of novelty and interest. Thus we will see
that there are charged edge states localized at $\pm \frac{L}{2}$
and that they are created by ``vertex" operators (or ``Wilson"
integrals). The Coulomb force between the charges comes out
correctly in this approach. We will also encounter fractionation of
charge by a mechanism first discovered in monopole theory by Witten
\cite{dyo,jack}. Just as for dyons, this phenomenon is related to the existence
of a ``$\theta$-angle" here as well. As a final result of this
section, we will find that there is a topological field theory which
describes these states. This transparent model in this manner nicely
illustrates important ideas found in more complex field theories.

Section 3 continues the preceding discussion to $2+1$ dimensional
spacetime with a disc or an annulus as the spatial slice. Just as in
Section 2, we find here also charge fractionation, vertex operators
and a topological field theory for edge states, the latter being a CS
theory for a generic MCS Lagrangian.  In this section we also summarize
the results found in \cite{higgs} for a gauged Higgs theory and briefly
discuss why this theory does not have edge states.

The discussion of the last two Sections tacitly assumes the
parameters $\lambda$ and $\mu$ mentioned above to be zero and is
therefore incomplete. We had previously discussed the origin of
these parameters from the characterization of certain boundary
conditions in the mathematical analysis, and also the quantization of
our Lagrangians for their generic values \cite{mcs,higgs}. In Section 4, we
summarize these boundary conditions for convenience while we
establish the physical meaning of $\lambda$ in Section 5. From this one sees
that it is possible
to vary $\lambda$ in an experimental arrangement. Now
physical properties are sensitive to their values and affected by
their variations. The energy spectrum for example is affected by
$\lambda$ and $\mu$ while edge states, existing in the MCS theory for
$\lambda=0$, get delocalized when $\lambda$ is
changed from this critical number. It seems for these reasons possible
to test our work experimentally. With this possibility in mind, we
also quote estimates for the dependence of the energy spectrum on
$\lambda$ in Section 6, using our work on the gauged Higgs
system for generic $\lambda$ (and $\mu =0$) \cite{higgs} as a guide.
This work incidentally not only quantizes the gauged
Higgs system for any $\lambda$ and $\mu$ but also argues that it does
not admit $U$ gauge for finite geometry and general $\lambda$ and
$\mu$.  The latter is established by showing that the gauged Higgs system in
the London limit is not equivalent to a massive vector meson theory for this
geometry.

Section 7 is the final one and concerns spatial manifolds with
boundaries in $3+1$ dimensions. It recalls previous work \cite{pao} on the $BF$
system showing
that edge states can occur in 3+1 dimensions as well. States
similar to the edge states of vortices remarked upon in Sections 3 and 4 are
shown to exist also for the 't Hooft-Polyakov monopoles. The possibility
of BC's depending on parameters like $\lambda$ and $\mu$, with
properties similar to those in 3+1 dimensions, is also shown. The
edge states spread out now too when $\lambda$ deviates from zero.
They should be observable by the dependence of their properties,
such as energies, on these parameters.

\sxn{Edge Observables in $1+1$ dimensions}

We here consider abelian Maxwell's theory on
$[-\frac{L}{2},\frac{L}{2}] \times {\bf R}^1$, $L$ being the spatial
length and ${\bf R}^1$ accounting for time. It is defined by the
Hamiltonian
\be
H = \frac{1}{2}\int_{-\frac{L}{2}}^{\frac{L}{2}} E(x)^{2} dx \; , \label{2.1}
\ee
the equal time commutators
\be
[A_{1} (x), A_{1} (y)]  =  [E(x),E(y)] = 0 \; , \nonumber
\ee
\be
[A_{1} (x), E(y) ]  =  i \delta (x-y) \label{2.2}
\ee
and the Gauss law constraint
\be
\frac{\partial E(x)}{\partial x} |\cdot \rangle = 0 \label{2.3}
\ee
on physical states $|\cdot \rangle$ . In these equations and in what
follows, we do not show the dependence of the fields on time. Also
$A_{1}$ here is the spatial component of the vector potential and $E$
is the electric field.

As emphasized elsewhere \cite{csb,camp}, the Gauss law (\ref{2.3}) should be
interpreted to mean
\be
G(\chi) |\cdot \rangle = 0 \; ,
\; G(\chi) := \int_{-\frac{L}{2}}^{\frac{L}{2}} d\chi E \label{2.4}
\ee
where the test function $\chi$ vanishes at the boundaries:
\be
\chi(x) |_{x=\pm \frac{L}{2}} = 0 \; . \label{2.5}
\ee

One way to justify the formulation of the quantum Gauss law using
(\ref{2.4},\ref{2.5}) is as follows.  A quantum field is an operator-valued
\underline{distribution}.  Classical equations involving its derivatives should
therefore be interpreted using operators obtained by smearing it with suitable
derivatives of test functions.  One such expression is in (\ref{2.4}).  But in
the classical limit, it gives the classical Gauss law only if (\ref{2.5}) is
true.  We thus justify (\ref{2.4},\ref{2.5}).

Alternative and conventional justifications use classical considerations about
the differentiability of functionals on the phase space \cite{csb,camp}.

We can now define two operators $Q_{\pm}$ corresponding to charges at
$\pm \frac{L}{2}$ as follows:
\begin{eqnarray}
Q_{\pm} &=& G(\Lambda_{\pm})  , \nonumber \\
\Lambda_{+}(x) |_{\frac{L}{2}} = 1 &,&
\Lambda_{+}(x) |_{-\frac{L}{2}} = 0 \; , \nonumber \\
\Lambda_{-}(x) |_{-\frac{L}{2}} = 1 &,&
\Lambda_{-}(x) |_{\frac{L}{2}} = 0 \; . \label{2.6}
\end{eqnarray}
The basis for the interpretation of $Q_{\pm}$ as charges at
$\pm \frac{L}{2}$ is discussed in \cite{camp}. Note that the action of
$Q_{\pm}$ on $|\cdot \rangle$ is independent of the value of
$\Lambda_{\pm}$ in the interior of our spatial interval because of
(\ref{2.4}). For this reason, we have not shown $\Lambda_{\pm}$ as an
argument of $Q_{\pm}$.

We emphasize that $Q_{\pm}$ are observables localized at the edges.
To see this, note that all observables localized in the interior of
our interval can be constructed from $E(x)$ ($|x| < \frac{L}{2}$)
and they commute with $Q_{\pm}$. So we can prepare states with
independent choices for the values of $E(x)(|x|<\frac{L}{2})$ and $Q_{\pm}$. In
addition, changing $\Lambda_{\pm}$ in the interior $|x| <
\frac{L}{2}$ does not change the action of $Q_{\pm}$ on
$|\cdot \rangle$ as observed previously. For these reasons, we can
think of them as edge observables.

We now show that
\be
(Q_{+} + Q_{-} ) |\cdot \rangle = 0 \label{2.7}
\ee
so that the charges at $\pm \frac{L}{2}$ add up to zero. The proof
follows from $Q_{+} + Q_{-} = G(\Lambda_{+} + \Lambda_{-})$ where
$\Lambda_{+} + \Lambda_{-}$ has the value 1 at both the edges. We
can therefore choose it to be the constant function with value 1
without altering $(Q_{+} + Q_{-}) |\cdot \rangle$ . But then
$Q(\Lambda_{+} + \Lambda_{-}) = 0$ and (\ref{2.7}) follows.

Next let us show that the spectrum of an edge charge is not
obliged to be quantized even if the group that is gauged is $U(1)$.
The mathematical reason behind this result is similar to the one
leading to the possibility of fractional charges for dyons pointed out by
Witten \cite{dyo}. The
group $\cal G$ of gauge transformations leaving the Hamiltonian
invariant are maps of $[-\frac{L}{2},\frac{L}{2}]$ to $U(1)$. It has
generators $G(\Lambda)$ with $\Lambda|_{\pm \frac{L}{2}}$ being
unrestricted. Of these, the subgroup ${\cal G}_{0}^{B}$ of gauge
transformations generated by $G(\chi)$, with $\chi$ obeying
(\ref{2.5}), leaves the physical states invariant. ${\cal G}_{0}^{B}$
is connected since $t\chi$ obeys (\ref{2.5}) for $0\leq t \leq 1$. [The
subscript of  ${\cal G}_{0}^{B}$ emphasizes this fact (while $B$
stands for boundary).] The effective symmetry group of quantum states
is hence ${\cal G}/{\cal G}_{0}^{B}$. The reason charge is not
quantized is because this group is not $U(1)$, but its universal
cover ${\bf R}^1$. The proof is as follows. We can choose $Q_{+}$ say
as its generator, remembering (\ref{2.7}). If $Q_{+}$ has a quantized spectrum,
then there
is a smallest period $\tau (>0)$ such that $e^{i\tau Q_{+}} |\cdot \rangle
= |\cdot \rangle$ or $e^{i\tau Q_{+}} \in {\cal G}_{0}^{B}$. ${\cal
G}_{0}^{B}$ being connected, it should therefore be possible to
deform  $e^{i\tau Q_{+}}$ to the identity staying within ${\cal
G}_{0}^{B}$. But that we can not do, $e^{it\tau Q_{+}}$ not being identity
on all states for $t<1$ by hypothesis. Charge is not therefore
quantized.

Just as for dyons, we can understand the possibility of fractional
charges from the existence of certain gauge transformations with
nontrivial topology. Thus consider the group ${\cal G}^B$ of all
gauge transformations $e^{i \Theta}$ which act as identities
at $\pm \frac{L}{2}$:
\be
e ^{i \Theta} |_{\pm \frac{L}{2}} = 1 \label{2.8}
\ee
or
\begin{eqnarray}
 \Theta (-\frac{L}{2}) & =& 0, \label{2.9} \\
\Theta (\frac{L}{2}) &=& 2  \pi N \; , \; N \in {\bf Z} \; . \label
{2.10}
\end{eqnarray}
Here (\ref{2.9}) is our choice of normalization for $\Theta$. From
(\ref{2.10}), one can see that ${\cal G}^B$ is the disjoint union of
${\cal G}^{B}_{N}$, with $\Theta$ for ${\cal G}^{B}_{N}$ fulfilling
(\ref{2.9},\ref{2.10}).  It is only
${\cal G}^{B}_{0}$ that is generated
by $G(\chi)$ and is required to act as identity on physical states.  If the
angle $\Theta$ associated with ${\cal G}^B_{N}$ is denoted by $\Theta _{N}$, it
follows that
the operator $T$ for the gauge transformation
$e^{i\Theta_{1}}$ can have eigenvalue $e^{i\theta}$ on a
physical state:
\be
T |\cdot \rangle = e^{i\theta} |\cdot \rangle \; . \label{2.11}
\ee

Note that the boundary condition (\ref{2.8}) allows us
to identify $x=\pm \frac{L}{2}$ and regard space as a circle $S^1$ when
discussing ${\cal G}^{B}$.
Thus ${\cal G}^B$ consists of maps of $S^1$ to $U(1)$ and these are
characterized by winding numbers. As these can be identified with
$N$, we see that (\ref{2.11}) is similar to the equation defining
$\theta$-states in familiar gauge theories like QCD \cite{call,book}.

Now
\be
T = e ^{iG(\Theta _{1})} = e ^{i2\pi Q_{+}} \; . \label{2.12}
\ee
It follows from (\ref{2.11}) that $|\cdot \rangle$ has charge $\theta /2\pi
\mbox{ mod } 1$, showing that it is in general fractional.
Fractional charge hence becomes possible here for topological reasons
similar to those for dyons.

There is even a topological term we can include in the action which
is associated with the $\theta$-states of (\ref{2.11}). It is the
analogue of the integral of $\theta Tr(F\wedge F)$ in QCD \cite{call}. It is
just
\be
\frac{\theta}{2\pi} \int d^{2}x (\partial_{0}A_{1} - \partial_{1}A_{0})
\label{2.13}
\ee and is familiar from studies of the Schwinger model \cite{schw}. [Here and
below, $x^{0}$ is time while $x^{1}$ is the same as the $x$ elsewhere in this
Section.]  Its
association to $\theta$-states is well known and will not be
repeated here.

Next consider the operator
\be
W = \int_{-\frac{L}{2}}^{\frac{L}{2}} dx A_{1} \; . \label {2.14}
\ee
It is invariant under the gauge transformations due to generators
$G(\chi)$ of ${\cal G}_{0}^{B}$:
\be
W \longrightarrow W + \int_{-\frac{L}{2}}^{\frac{L}{2}} d\chi = W \;
. \label{2.15}
\ee
It is hence an observable. It is conjugate to $Q_\pm$:
\be
[W,Q_{\pm}] = \pm i \; . \label{2.16}
\ee
The operator
\be
V(e) = e^{ieW} \label{2.17}
\ee
therefore creates charges $\pm e$ at $\pm \frac{L}{2}$:
\be
Q_{\pm} V(e) |q\rangle = (q+e) V(e) |q\rangle \mbox{ if } Q_{\pm}
|q\rangle = \pm q \, |q\rangle \; . \label{2.18}
\ee
$V(e)$ is the analogue of the vertex operator in conformal field
theories (CFT's) \cite{vo} and will be called by the same name here.

This vertex operator shifts the electric field by a constant:
\be
V(e)^{-1} E(x) V(e) = E(x) + e \; . \label{2.19}
\ee
Hence if $|0\rangle$ is the physical state with zero electric field and energy,
then the energy of $V(e)|0\rangle$ is $\frac{1}{2}e^{2}L$, its electric field
being $e$:
\be
H V(e) |0\rangle = (\frac{1}{2}e^2 L) V(e) |0\rangle \; . \label{2.20}
\ee
In other words, the vertex operator approach gives the correct linear
potential between the charges.

Since $V(e)$ does not commute with $E(x)$ for $|x| < \frac{L}{2}$, it
cannot be thought of as localized at the edge. That is hence also the case
for the excitations $V(e)$ creates by acting for
instance on $|0\rangle$. Now as $Q_\pm$ fails to commute with $V(e)$, the
view that $Q_\pm$ is localized at the edge requires comment. In CFT,
$V(e)$ is not regarded as an observable, rather the algebra of
observables is regarded as being generated by $E(x)$ and $Q_\pm$. In
this point of view, which we adopt, $V(e)$ intertwines different
representations of this algebra, while $Q_\pm$, which commutes with
$E(x)(|x|<\frac{L}{2})$, can be thought as localized at the edge.

There is also a description of edge observables using a topological ``$BF$"
theory involving no metric. $B$ here is a scalar field, we write it
as $E$ as it will later get identified with the electric field. As for
$F$, it is just the electromagnetic field tensor $dA$, $A$ being
$A_0 dx^0 + A_1 dx^1$. The action is
\be
S = \int E dA = \int dx^0 dx^1 [ E \partial_0 A_1 - E \partial_1 A_0]
\; . \label{2.21}
\ee
The first term gives (\ref{2.2}) while the second gives (\ref{2.3}).
It thus correctly reproduces the algebra of observables of the Maxwell
theory. Of course, just as all topological field theories, it gives
zero for the Hamiltonian.

Summarizing, observables localized at the edges are charges, the
total charge of both the edges being zero and each edge charge having the
option of being
fractional. Wilson integrals or vertex operators can create them and
also correctly reproduce their Coulomb attraction. But their kinetic
energies have to be added by hand to get a complete description of
their dynamics.

\sxn{Edge Observables in $2+1$ Dimensions}

In this section, we work with a round disk $D$ with a round hole $H$
in the middle, so that the spatial slice is the annulus $D \backslash
H$.

We begin by recollecting old results for CS and MCS actions
\cite{cs,witt,csb,mcs,pao}.  We then move on to discuss the Higgs model.
It may be noted that the material here corresponds to the case
$\lambda = 0$ in ref. \cite{mcs} and $\lambda=\mu=0$ in ref. \cite{higgs}.

\subsxn{The CS Observables at the Edge}

Chern-Simons Lagrangians were studied  in \cite{cs,witt,csb}. They define
affine
Lie algebras and can be interpreted in terms of two-dimensional massless
scalar fields. There are no excitations whatsoever in the interior of
$D \backslash H$ in these models.

The Lagrangian and the non-vanishing equal time commutators for the $U(1)$ CS
model are
\be
L = \frac{k}{4\pi} \int_{D\backslash H} d^{2}x\epsilon^{\mu\nu\rho} A_{\mu}
\partial_{\nu} A_{\rho} \label{3.1}
\ee
\be
[A_{1}(x),A_{2}(y)] = i\frac{2\pi}{k}\delta(x-y) \label{3.2} \; .
\ee

The Gauss law operator for $L$ is
\begin{eqnarray}
&&G(\chi) = \frac{k}{2\pi}\int_{D\backslash H} d\chi A \; , \; \chi|_{\partial
D} =
\chi|_{\partial H} = 0 \; ,\nonumber\\
&&A:=A_{\mu}dx^{\mu}. \label{3.3}
\end{eqnarray}
[Here and in what follows, we omit the wedge symbol between differential
forms.] It annihilates physical states $|\cdot \rangle$:
\be
G(\chi) |\cdot \rangle = 0 \; ,\label{3.4}
\ee
and generates a group ${\cal G}_0^B$.

The charge operators $Q_\pm$ for the outer and inner edges $\partial
D$ and $\partial H$ are given by
\be
Q_{\pm} = G(\Lambda_{\pm}) \; , \nonumber
\ee
\be
\Lambda_{\pm} |_{|\vec{x}|=R_{\pm}} = 1 \; , \;
\Lambda_{\pm} |_{|\vec{x}|=R_{\mp}} = 0 \; , \label{3.5}
\ee
$\partial D$ having the radius $R_{+}$ and $\partial H$ the radius
$R_{-}$. They commute.  We can show as previously that
\be
(Q_{+} + Q_{-}) |\cdot \rangle = 0 \; , \label{3.6}
\ee
that is that $D\backslash H$ has zero net charge.

The spectrum of $Q_{+}$ is not quantized here just as it previously
was not. The proof is the same as before: $e^{i\Lambda Q_{+}}$ is
not in ${\cal G}_0 ^B$ for any nonzero $\Lambda$.

We will omit writing the generators of the affine Lie algebras \cite{witt,csb}
at the two edges.

As in Section 2, or as for dyons, there are also nonzero winding number gauge
transformations which can be regarded as the cause of charge fractionation.
These transformations become identity at $\partial H$ and $\partial D$ and wind
around $U(1)$ an integral number of times as the radial coordinate increases
from a point on $\partial H$ to a point on $\partial D$ \cite{unp}.

There is even an analogue of (\ref{2.13}).  If $\phi$ is the azimuthal angle in
the annulus, it is \cite{unp}
\be
\mbox{constant }\times\int d\phi dA . \label{tfa}
\ee

The vertex operator $V$ is essentially the parallel transport operator
\be
\exp ieW \equiv \exp ie \int_{L} A \; , \label{3.7}
\ee
the line $L$ starting at a point $P_{-}$ on $\partial H$ and ending
at a point $P_{+}$ on $\partial D$. As discussed elsewhere \cite{csb},
$V$ actually requires regularization by normal ordering so that its
correct form is
\be
V = : \, e^{ieW} \, : \; . \label{3.8}
\ee
It creates charges $\pm e$ at $P_{\pm}$ when applied to a state and
is basically the Fubini-Veneziano vertex operator.

The Hamiltonian for the CS action being zero, we can not now calculate an
interesting Coulomb energy for charges unlike in Section 2.

\subsxn{The MCS Observables at the Edge}

The MCS Lagrangian is
\begin{eqnarray}
L & = & \int_{D \backslash H} d^{2}x {\cal L} \nonumber \\
{\cal L} & = & -\frac{1}{4e^{2}} \,
F_{\mu \nu}F^{\mu \nu}+\frac{k}{4\pi}\,
\epsilon ^{\mu \nu \lambda}A_{\mu}\partial _{\nu}A_{\lambda} \; .
 \label{3.9}
\end{eqnarray}
The Hamiltonian and Gauss law for (\ref{3.9}) are
\begin{eqnarray}
&~& H  =  \int_{D \backslash H} d^{2}x \: {\cal H}  \nonumber \\
&~&{\cal H}  =  \frac{e^{2}}{2} [(\Pi _{i} +\frac{k}{4\pi}\epsilon
_{ij}A_{j})^{2} + \frac{1}{e^{4}}(\epsilon _{ij}
\partial _{i}A_{j})^{2}] \; \nonumber\\
&~& G(\chi) |\cdot \rangle = 0 \mbox{ for } \chi|_{\partial
D,\partial H} = 0 \; , \label{3.10}
\end{eqnarray}
where
\be
G(\chi) = -\int_{D \backslash H} d^{2}x\,\partial
_{i}\chi ^{(0)}[\Pi _{i}-\frac{k}{4\pi}\epsilon_{ij}A_{j}]
\label{3.11}
\ee
and $|\cdot \rangle$ is any physical state.

The edge observables $Q (\Lambda_{\pm})$ for $\partial D$ and
$\partial H$ are obtained as before from $G$ by changing the boundary
conditions on $\chi$. They are
\be
Q(\Lambda_{\pm}) = -\int_{D \backslash H} d^{2}x\,\partial
_{i}\Lambda_{\pm} [\Pi _{i}-\frac{k}{4\pi}\epsilon_{ij}A_{j}]
\; , \; \Lambda _{+}|_{\partial H} = \Lambda _{-}|_{\partial D} = 0 \; .
\label{3.12}
\ee
They are the generators of two affine Lie groups $\tilde{L}U(1)$ and
have the commutators
\be
[Q(\Lambda_{\epsilon}),Q(\Lambda '_{\epsilon'})] = -i\frac{k}{2\pi}\delta_
{\epsilon\epsilon'}\int _{D\backslash H}d\Lambda _{\epsilon}d\Lambda
'_{\epsilon '} \; . \label{3.13}
\ee
As argued previously, since the action of $Q(\Lambda_{\epsilon})$ on
$|\cdot \rangle$ depends only on the boundary value of
$\Lambda_{\epsilon}$, and since it commutes with observables
localized within $D \backslash H$, it can be regarded as localized at
the edge.

The charges $Q_{\pm}$ at $\partial D$ and $\partial H$ are special
cases of $Q(\Lambda_{\pm})$:
\begin{eqnarray}
Q_{+} = Q(\Lambda_{+}) & \mbox{with} & \Lambda_{+}(x)|_{x\in\partial
D} = 1 \; , \nonumber \\
Q_{-} = Q(\Lambda_{-}) & \mbox{with} & \Lambda_{-}(x)|_{x\in\partial
H} = 1 \; . \label{3.14}
\end{eqnarray}
We have, as usual,
\be
(Q_{+} + Q_{-}) |\cdot \rangle = 0 \label{3.15}
\ee
which means that $D\backslash H$ has zero total charge.

For $k\neq 0$, the topological field theory describing the edge states
associated with (\ref{3.13}) is just the CS theory, as one can readily show
using the results of \cite{witt,csb}.

In the present case too, just as for the $CS$ case, the spectra of
$Q_{\pm}$ are not obliged to be quantized for reasons similar to those in
Section 3.1.  The topological term analogous to (\ref{2.13}) is the same as in
that Section, namely (\ref{tfa}).  Also there is a vertex
operator $V$ for the creation of charges similar to (\ref{3.8}) and we can as
in Section 2
calculate the Coulomb energy for the charges created by $V$.

The $k\rightarrow 0$ limit of the MCS system is not smooth \cite{mcs}.
The choice $k=0$ gives the pure Maxwell theory, and that too can have
edge observables, but with properties different from the MCS edge
observables.

\subsxn{Edge Observables and the Higgs Model}

We next summarise the work of ref \cite{higgs} showing that there are
no edge observables in the Higgs model.

Consider the $U(1)$ Higgs model on $D\backslash H$ with the modulus of the
Higgs field frozen
to its vacuum value.  Its only degree of freedom left is then its phase
$e^{iq\psi}$ where $q(\neq 0)$ is the charge of the Higg's field.  In this so
called London limit, the Lagrangian reads
\be
L = \int_{D \backslash H} d^{2}x\left\{ -\frac{1}{4 e^2} F_{\m\n} F^{\m\n} -
\frac{m^2_{H}}{2} (\pa_\m \psi - A_\m) (\pa^\m \psi - A^\m) \right\} ,
\label{3.16}
\ee
where the parameter $q$ has been absorbed along with the vacuum value of the
Higg's field (and a constant accounting for the thickness of the disk)
into an effective parameter $m_{H}$ (the vector meson mass being equal to
$em_{H}$).

It gives rise to the Gauss law
\be
G(\chi) = \int _{D\backslash H}d^{2}x (\chi\Pi + \frac{1}{e^{2}}\partial
_{i}\chi E _{i})
, \; \chi|_{\pa D,\pa H} =0 \; , \label{3.17}
\ee
annihilating the physical states $|\cdot \rangle$ ,
\be
G(\chi) |\cdot \rangle = 0    \; ,      \label{3.18}
\ee
$\P$ and $E_i/e^{2}$ being respectively the momenta conjugate to $\psi$ and
$A_i$.

As before, we might want to consider the observables
\begin{eqnarray}
Q(\Lambda _{\pm}) & = & \int _{D\backslash H}d^{2}x(\Lambda _{\pm}\P  +
\frac{1}{e^{2}}\partial _{i}\Lambda _{\pm} E_{i}), \label{edhi}\\
\Lambda _{+} |_{\partial H} & = 0 = & \Lambda _{-}|_{\partial D} \;,
\nonumber\\
\Lambda _{+} |_{\partial D} &\mbox{and}& \Lambda _{-}|_{\partial H}\; \neq \; 0
,\label{3.19}
\end{eqnarray}
and argue that $Q(\Lambda _{\pm})$ define edge states because of (\ref{3.18})
and because they commute with observables localised in the interior of
$D\backslash H$.  If the excitations described by (\ref{3.16}) correspond to
the fluctuations about a vortex solution (where the singularity in $\psi$ is
wholly confined within the hole), then these edge observables will be
associated with edge states associated to a vortex.

There is however a major difference between this expression for $Q(\Lambda
_{\pm})$ and the corresponding one in the MCS theory.
In the first term of (\ref{edhi}), it is the function
$\L_{\pm}$ that smears the momentum $\P$. But we can always choose
to expand a scalar function on a disc (or a disc with an annulus) in a basis of
functions all of which vanish on both the boundaries. That is,
$\L_{\pm}$ can always be written as a (possibly infinite) sum of functions
$\chi_{n_1 n_2 \cdots}$, for a suitable set of indices $\{n_1 n_2 \cdots\}$,
where $\chi_{n_1 n_2 \cdots}|_{\pa D,\pa H} = 0$.  This is so despite the
second line of
(\ref{3.19}), convergence in the expansion being in the $L^{2}$-sense.  Hence,
by virtue of
(\ref{3.18}), $Q(\Lambda_{\pm})$ annihilates any physical state $|\cdot
\rangle$ and represents
the trivial observable. Edge states are thus absent in the Higgs system [at
least with this treatment of $Q(\Lambda _{\pm})$].

That this is not the case in the MCS theory
can be seen by noticing that the fields
are smeared out by derivatives $\pa_i \L_{\pm}$ of the test functions
$\L_{\pm}$ in (\ref{3.12}). Now (as we have rigorously proved in \cite{mcs})
$\pa_i \L_{\pm}$ can, and actually has to, be expanded as a sum of functions
that do not necessarily vanish on the boundary.  Hence $Q_{\pm}$ of
(\ref{3.14})
do not annihilate the
physical states.  In fact, it is impossible to expand $d\L_{+}\;(d\L_{-})$ in
terms of exterior derivatives of functions that vanish at $\partial
D\;(\partial H)$ because $d\L_{+}\;(d\L_{-})$ is always orthogonal to these
forms. The observables
$Q_{\pm}$ are therefore non-trivial.

\sxn{The Boundary Conditions}

In this Section, we work in 2+1 dimensions and start by stating the BC's for
the MCS Lagrangian and the
Maxwell and CS cases which are its limiting forms.  We next do the same for the
Higgs problem.  From here onwards in this paper, we assume for simplicity that
there is no hole so that our spatial manifold is $D$.  There is of course no
difficulty at all in restoring the hole.

\subsxn{The MCS Lagrangian and its Limiting Forms}

Quantisation in $D$ involves specification of BC's on $\partial D$.  In this
case, as shown elsewhere \cite{mcs}, there exists a one-parameter family of
BC's compatible with locality and
nonnegativity of the Hamiltonian for the MCS Lagrangian and its limiting forms.
They
are parametrised by a number $\lambda \geq 0$.  For each $\lambda$, they
specify the following domain ${\cal D}_{\lambda}$ for a certain Laplace like
operator relevant for quantisation:
\be
{\cal D}_{\lambda} = \{ A| *dA|_{\partial D} = -\lambda A_{\theta}|_{\partial
D}
; \lambda \geq 0 \}. \label{4.1}
\ee
Here $*$ is Hodge star.  Throughout this paper, we use the Hodge star only on
a spatial slice and so we will need its definition only on a Euclidean
manifold.  The Hodge star of a $p$-form $\alpha$ on an
$n$-dimensional Euclidean manifold with flat metric is an
$n-p$ -form $*\alpha$ defined as follows:
\be
\label{hodge*}
(*\alpha )_{i_{1}i_{2}\ldots i_{n-p}}= \frac{1}{p!}\epsilon _{i_{1}i_{2}\ldots
i_{n-p}j_{1}j_{2}\ldots j_{p}}(\alpha )_{j_{1}j_{2}\ldots j_{p}}.
\ee
Here $\epsilon $ is the totally antisymmetric Levi-Civita form while the
components
of the forms are taken along an orthonormal basis.  The components of an
$r$-form $\omega $ are themselves defined via the relation
\be
\label{defcom}
\omega =\frac{1}{r!}\omega _{i_{i}\ldots i_{r}}dx^{i_{1}}\ldots dx^{i_{r}}
\ee
where the $dx^{i_{k}}$'s are the one-forms along the orthonormal coordinate
basis.  Let us also note our
definitions of the wedge product, and the Hodge star of the exterior
derivative, in component form:
\begin{eqnarray}
&&(\a ^{(p)}\wedge \b ^{(q)})_{i_{1}\ldots i_{p}i_{p+1}\ldots
i_{p+q}}=\frac{1}{p!q!}\epsilon _{i_{1}\ldots i_{p+q}}\a _{i_{1}\ldots i_{p}}\b
_{i_{p+1}\ldots i_{p+q}}, \label{defw}\\
&&(*d\a ^{(p)})_{i_{1}\ldots i_{n-p-1}}=\frac{1}{p!}\epsilon _{i_{1}\ldots
i_{n-p-1}i_{n-p}\ldots i_{n}}\partial _{i_{n-p}}\a _{i_{n-p+1}\ldots i_{n}}.
\label{defd}
\end{eqnarray}
Here and below $d$
on a form always refers to its exterior derivative restricted to the spatial
manifold.

Also the $A_{\theta}$ in (\ref{4.1}) has the familiar meaning:
\be
A_{\theta} = A_{i}
\frac{1}{r}\frac{\partial x^{i}}{\partial \theta}(r,\theta ). \label{4.2}
\ee
Here $r,\; \theta$ are polar coordinates on $D$ with $r=R$ corresponding to
$\partial D$.

Let us also note here that (as explained at length in \cite{mcs}) the edge
observables discussed in Section 3.2 exist only when the
parameter $\lambda$ defined above vanishes.

\subsxn{The Higgs Model in the London Limit}

BC's compatible with locality and energy nonnegativity for (\ref{3.16}) have
been derived in \cite{higgs}.  They are parametrised by two nonnegative numbers
$\lambda , \mu$, the pair $\lambda,\mu$ specifying a domain ${\cal D}_{\lambda,
\mu}$ for a certain
second order operator.  The definition of ${\cal D}_{\lambda ,\mu}$ is
\be
{\cal D}_{\lambda, \mu} = \{ (\psi,A) \;| \; *dA|_{\partial D} = -\lambda
A_{\theta}|_{\partial D}\; , \; \psi|_{\partial D}=-\m (\pa_{r}\psi - A_r )
|_{\partial D} \; ; \; \lambda
,\mu \in \, I \!\! R \; ,\; \l , \m \geq 0 \}  \; .\label{4.3}
\ee

The following point is worthy of note.  On $D$, with the magnitude of the Higgs
field frozen to a constant, $e^{iq\psi}|_{\partial D}$ can have only zero
winding number since $e^{iq\psi}$ will not exist as a smooth function on all of
$D$ in the contrary case.  Hence $\psi |_{\partial D}$ is a well defined
(single-valued)
function.  If this assumption is relaxed by for example restoring the hole,
$e^{iq\psi}|_{\partial D}$ may have a non-zero winding number, say $N\in Z$.
We must then replace $\psi$ in (\ref{4.3}) by $\Delta \psi = \psi -
2\pi N/q$.  This will be the appropriate scenario for the description of
quantum excitations about the classical vortex solutions.

\sxn{Physical Interpretation of Boundary Conditions}

\subsxn{The MCS Model}

We assume that $D$ is surrounded by a superconductor with penetration depth
$1/m$, and that the situation
in the exterior near the edge is static, or in
other words that time scales for dynamics near the edge outside the disc are
long
compared to typical time scales of our interest.

The order parameter for a superconductor is a complex Higgs field.  It is a
good approximation to take its modulus to be fixed at the vacuum value for
processes involving moderate energies.  Let us make this assumption and
approximate the Higgs field by the field $e^{iq\psi}$, $\psi$ being real
valued.
The field equations in the exterior near $\partial D$ that arise from the
variation of (\ref{3.16}) lead to
\begin{eqnarray}
d*d(A-d\psi ) - e^{2}m^{2}_{H}*(A-d\psi ) & = & 0,\label{5.1} \\
d*(A-d\psi ) & = & 0, \label{5.2}
\end{eqnarray}
(\ref{5.2}) following from (\ref{5.1}) on applying $d$.  Here we have assumed
that the fields are static and have used the
definitions given in equations (\ref{hodge*}) to (\ref{defd}).  Also we have
not written the equations involving $A_{0}-\partial _{0}\psi$.

Now (\ref{5.1}) and
(\ref{5.2}) imply that $A-d\psi$ satisfies Laplace's equation:
\be
\vec{\nabla} ^{2}(A-d\psi )= e^{2}m^{2}_{H}(A-d\psi ). \label{lap}
\ee
Assuming variations along azimuthal direction to be suppressed compared to
radial fluctuations, this equation tells us that $A-d\psi$ decays like
$e^{-em_{H}r}$ outside $D$, $r$ being the distance from $D$.  Thus the inverse
penetration depth $m$ is equal to $em_{H}$.

We will now examine the conditions on $A$ and $\psi$ near $\partial D$
following from (\ref{5.1},\ref{5.2}).  Their compatibility with (\ref{4.1})
will establish the claimed result that $1/\lambda$ is proportional to the
penetration depth $1/m$, and will also constrain $\psi$.

Let $P$ be a point on $\partial D$, and $Q$ a point at a distance $\Delta$ from
$P$ in the radial direction.  On integrating radially from $P$ to $Q$,
(\ref{5.1}) gives
\be
(*dA)|_{Q}-(*dA)|_{P}-m^{2} \int _{P}^{Q} *(A - d\psi )=0. \label{5.3}
\ee

Now
\begin{eqnarray}
\int _{P}^{Q} *(A -d\psi ) = \int _{P}^{Q} \varepsilon _{ij}(A_{j} -\partial
_{j}\psi )\frac{\partial x^{i}}{\partial r}\, dr & = & \int _{P}^{Q}(A_{\theta}
-\frac{1}{r}\partial _{\theta}\psi )\, dr, \label{5.4} \\
A_{\theta} - \frac{1}{r}\partial _{\theta}\psi & = & (A_{i} -\partial _{i}\psi
)\frac{1}{r}\frac{\partial x^{i}}{\partial \theta} ,\label{5.5}
\end{eqnarray}
$r$ and $\theta$ being polar coordinates with $r=0$ being the center of the
disc.  With $m\Delta >>1$, the right hand side of (\ref{5.4}) is approximately
$(A_{\theta} -\frac{1}{r}\partial _{\theta}\psi )(P) /m$.  For example, if $R$
is the radius of $D$, and we write $(A_{\theta} -\frac{1}{r}\partial _{\theta}
\psi ) (x) = (A_{\theta} - \frac{1}{r}\partial _{\theta}
\psi )(P) e^{-m(r-R)}$, it is $\frac{1}{m} (A_{\theta} -\frac{1}{R}\partial
_{\theta}\psi )|_{P}$. Also under the same condition, we can assume that
$(*dA)_{Q}(= *d(A -d\psi )|_{Q})$ is approximately zero.  So
\be
(*dA)|_{P}\cong -m(A_{\theta}-\frac{1}{R}\partial _{\theta}\psi )|_{P} \; .
\label{5.6}
\ee
Comparison with (\ref{4.1}) then gives
\begin{eqnarray}
& &\lambda =  m, \label{5.7} \\
& &\partial _{\theta} \psi |_{\partial D} =  0 \; .\label{5.8}
\end{eqnarray}
The interpretation of $1/\lambda$ in terms of the penetration depth is thus
established.
Equation (\ref{5.7}) being in the nature of an estimate, it is best regarded as
the statement of proportionality of $\lambda$ and $m$.

Equation (\ref{5.7}) requires that $\lambda$ is positive.  It is significant
that the
same sign can be derived by requiring that the Hamiltonian is non-negative.

Equation (\ref{5.8}) is a condition on the static superconductor surrounding
$D$.  It is
to be used as a boundary condition when solving (\ref{5.1}).  Although it is
not invariant under gauge transformations which do not act as identity on
$\partial D$, it is invariant under gauge transformations generated by the
Gauss law, the latter becoming identity on $\partial D$.

As (\ref{5.2}) is a consequence of (\ref{5.1}), there is no need to examine it
separately here.

\subsxn{The Higgs Model}

As before, we assume that $D$ is surrounded by a superconductor governed by the
equations (\ref{5.1}) and (\ref{5.2}).  In the interior of $D$ too, we have a
superconductor described by the fields $A,\tilde{\psi}$ (the latter decorated
with a tilde to distinguish it from the phase $\psi$ of the surrounding
superconducting medium) and the boundary conditions
(\ref{4.3}) with $\tilde{\psi}$ replacing $\psi$.

Assuming the exterior of $D$ to be static, we get (\ref{5.6}) once more.  Its
comparison with (\ref{4.3}) shows that (\ref{5.7}) and (\ref{5.8}) are still
valid, and that $1/\lambda$ is proportional to the penetration depth in the
exterior of $D$.

An interpretation for $\mu$ along similar lines is lacking because we do not
have the analog of a ``Meissner effect'' for the phase of the Higgs field.
Note however that if the Higgs model is to serve as an
approximation for the MCS model, then the parameter $\mu$ must be zero, as will
be shown in Section 6.

\sxn{The Energy Spectrum and its Dependence on $\lambda$}

We have already remarked that physical quantities are sensitive to the BC's
obeyed by the fields.
In this section, we will study the energy spectrum of the MCS and Higgs
theories in a more quantitative way. In particular we will study how the
spectrum changes when we vary the parameter $\l$. [For reasons explained below
we will fix $\m$ to be zero.]

Our discussion of the MCS theory in \cite{mcs} was limited by the fact that,
for
BC's charcaterized by a nonzero value of $\l$, we were not able to diagonalize
the Hamiltonian in (\ref{3.10}) and could not therefore study its spectrum
except when $\l=0$.

On the other hand, in \cite{higgs} we have shown that the Higgs Hamiltonian
for the Lagrangian (\ref{3.16}) in the
London limit can be exactly diagonalized for any value of the
parameters $\l ,\m$ characterizing its BC's. We have also noticed that for $\m
=0$, there is a one-to-one correspondence between the modes
present in the Hamiltonian of the Higgs theory and the ones in the Hamiltonian
of the MCS theory provided they both have the same value of $\l$. Indeed, both
these theories
describe massive vector mesons (with masses $em_{H}$ and $\frac{e^{2}k}{2\p}$
respectively), the only major difference being the existence, for $\lambda =0$,
of edge states in the latter and not in the former.

We might therefore argue that
the Higgs Hamiltonian will serve us as a guide in the study of the MCS theory
as
well. The following results about the energy spectrum will be exact for the
Higgs model for any value of $\l$.  But they will be exact for the MCS theory
only for $\l=0$, that is when
we are able to diagonalize its Hamiltonian exactly. For the MCS theory with $\l
>0$, they represent only an approximation, which is probably reasonable if $\l$
deviates from zero only slightly.

In \cite{higgs}, we have second quantized the Higgs system in the London limit
and shown that its Hamiltonian can be written (when $\m=\l=0$) as
\be
H = \O_{nm}^{(\a )} a_{nm}^{(\a )\dag} a_{nm}^{(\a )} +
\O_{nm}^{(\b )} a_{nm}^{(\b )\dag} a_{nm}^{(\b )} +
\O_{n}^{(h)} b_n^{\dag} b_n \; ,
\label{6.1}
\ee
where the $a^{(j)}_{nm},a^{(j)\dag}_{nm}$ ($j=\a ,\b$ ) and the
$b_{n},b_{n}^{\dag}$ are annihilation-creation operators for the three kinds of
modes that appear in the mode expansion of $(A(x), \psi (x))$ when $\mu
=\lambda =0$.  Their non-zero commutators are
\bea
\left[ a_{nm}^{(j)},a_{n'm'}^{(k)\dag} \right] &=& i\d_{jk} \d_{nn'} \d_{mm'},
\nonumber\\
\left[ b_n , b_{n'}^{\dag} \right] &=& i \d_{nn'} \; .
\label{6.2}
\eea
When $\l\neq 0$ (but $\m$ still equal to 0), the Hamiltonian takes an almost
identical form save for the modes $b_{n},\; b_{n}^{\dag}$ being absent, the
latter having been deformed into a $\b$-mode as we will see shortly.

The energies
$\O_{nm}^{(\a )},\; \O_{nm}^{(\b )},\; \O_{n}^{(h)}$
corresponding to the particle creation operators $a_{nm}^{(\a )\dag},
a_{nm}^{(\b )\dag},b_n ^{\dag}$ are respectively
\be
\O_{nm}^{(\a )}= \sqrt{\a^2_{nm} +  e^{2}m^2_{H}} \;,\;
\O_{nm}^{(\b )}= \sqrt{\b_{nm}^2 +  e^{2}m^2_{H}} \;,\;
\O_{n}^{(h)} = e m_{H} \; , \label{6.3}
\ee
where the real numbers $\a_{nm},\b_{nm}$ are fixed by the following BC's for
$\m=0$ and arbitrary $\l $:
\bea
J_n (\a_{nm}R) &=& 0 \; ,\; \a_{nm}\neq 0 ,\nonumber \\
\b_{nm} J_n (\b_{nm}R) &=& \l \left[ \frac{d}{d(\b_{nm} r)}
J_n(\b_{nm} r) \right] _{r=R} \; .\label{6.4}
\eea
Here, $J_n (x)$ is the real Bessel function of order $n$.
Incidentally, we see from (\ref{6.3}) that for the case $\m =\l =0$ (the case
for which the modes $b_{n},b_{n}^{\dag}$ exist), there is an infinite
degeneracy
of the spectrum at precisely the energy equal to the Higg's mass.
Also since only the equation defining the $\b_{nm}$ depends on $\l$,
$\O_{nm}^{(\a )}$, unlike $\O_{nm}^{(\b )}$, will not change when $\l$ is
changed.

Figures 1 to 3 give plots of
\be
G_n (\o R) = \frac{\o R J_n (\o R)}{J_n '(\o R)} \; \; , \; \;
J_n '(\o R) := \left[ \frac{d}{d(\o r)} J_n(\o r) \right] _{r=R}
\label{6.5}
\ee
versus $\omega R$ for $n=0,1$ and 10 respectively.  We can identify lines of
constant $\lambda R$ in these figures with lines parallel to the abscissas, the
$\omega R$-coordinates of their intersections with the $G_{n}(\omega R)$ versus
$\omega R$ giving $\beta _{nm}$.  The intersections of the graphs of the
functions with the abscissas at non-zero $\o$ give instead $\alpha _{nm}$.

\begin{figure}[p]
\begin{center}\mbox{\psannotate{\psbox{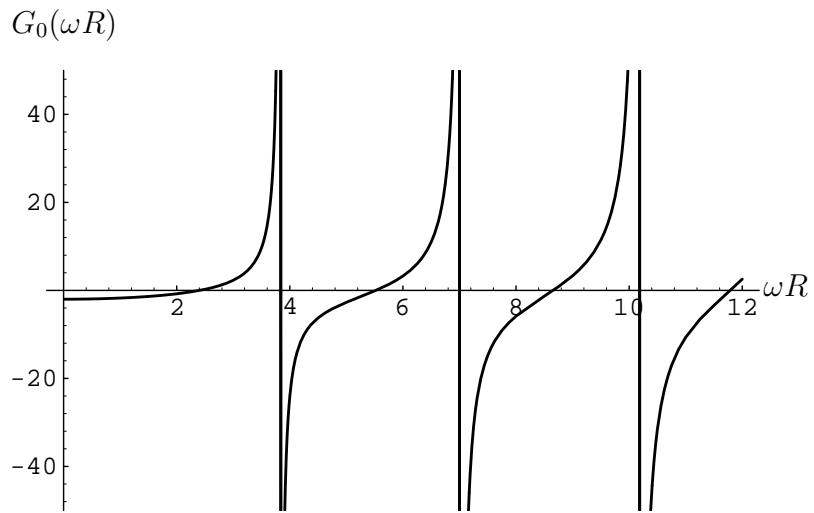}}
{\at(0\pscm;6.5\pscm)
{$G_{0}(\omega R)$}\at(10\pscm;3\pscm){$\omega R$}}}\end{center}
\caption{This figure gives the plot of $G_{0}(\omega R)$ vs $\omega R$.}
\end{figure}
\begin{figure}[p]
\begin{center}\mbox{\psannotate{\psbox{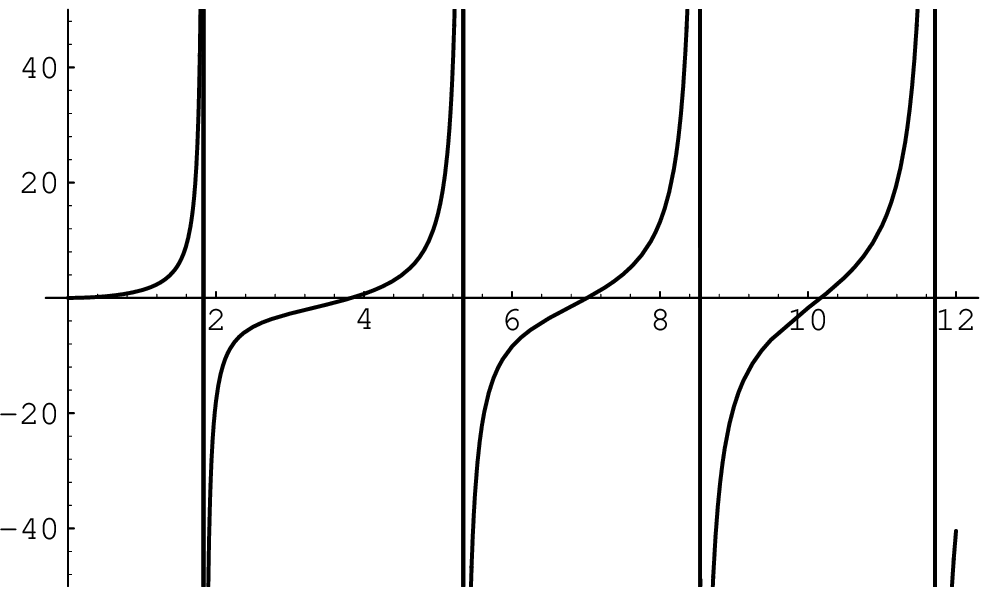}}
{\at(0\pscm;6.5\pscm)
{$G_{1}(\omega R)$}\at(10\pscm;3\pscm){$\omega R$}}}\end{center}
\caption{This figure gives the plot of $G_{1}(\omega R)$ vs $\omega R$.}
\end{figure}
\begin{figure}[hbt]
\begin{center}\mbox{\psannotate{\psbox{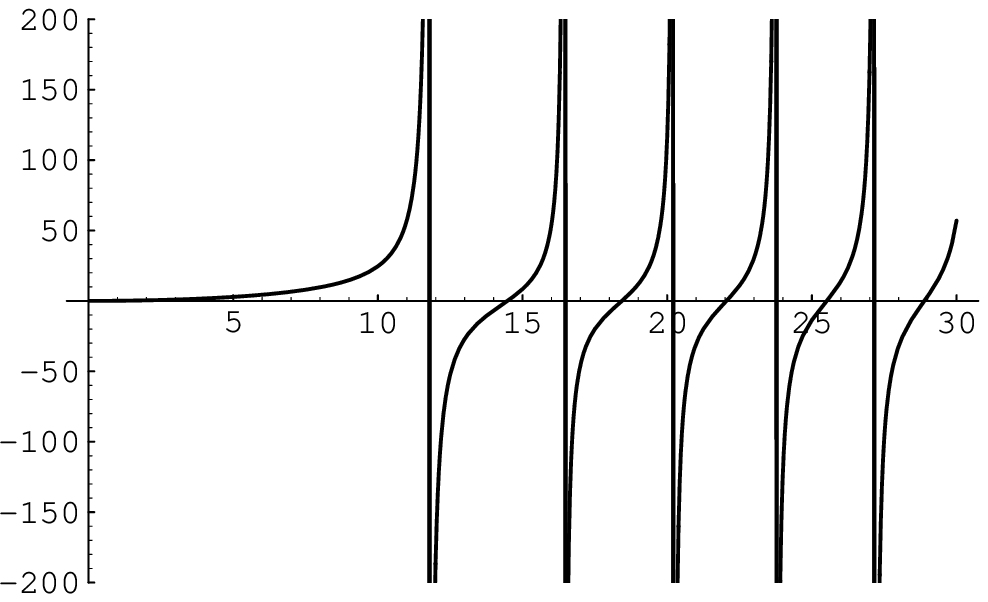}}
{\at(0\pscm;6.5\pscm)
{$G_{10}(\omega R)$}\at(10\pscm;3\pscm){$\omega R$}}}\end{center}
\caption{This figure gives the plot of $G_{10}(\omega R)$ vs $\omega R$.}
\end{figure}

We can now numerically calculate the dependence on $\lambda $ of some
of the energies $\O_{nm}^{(\a )},\;\Omega _{nm}^{(\beta )}$ in (\ref{6.3}).

Firstly, the
$\alpha _{nm}R$, which are the same as $\beta _{nm}R$ when $\lambda =0$,
remain unchanged for all $\l$.  Their values for the first few
allowed $m$'s when $n=0,1$ or 10 are approximately:
\begin{eqnarray}
&&\mbox{for }n=0 \;\; : \;\; 2.4,\;\;\; 5.5,\;\;\; 8.6,\ldots \nonumber\\
&&\mbox{for }n=1 \;\; : \;\; 0,\;\;\; 3.8,\;\;\; 7.0,\ldots \nonumber\\
&&\mbox{for }n=10 \;\; : \;\; 0,\;\;\; 14.5,\;\;\; 18.4,\ldots \label{j0root}
\end{eqnarray}
Here $m=1,2$ and 3 respectively in each of the rows.

Similarly, the $\beta _{nm}R$ for $\lambda$ very large (close to infinity so
that the relevant points correspond to the roots of $J'_{n}(\omega R)$) again
for the first few $m$'s when $n=0,1$ or 10 are approximately:
\begin{eqnarray}
&&\mbox{for }n=0 \;\; :\;\; 3.8,\;\;\; 7.0,\;\;\; 10.2,\ldots \nonumber\\
&&\mbox{for }n=1 \;\; :\;\; 1.8,\;\;\; 5.3,\;\;\; 8.5,\ldots \nonumber\\
&&\mbox{for }n=10 \;\; :\;\; 11.8,\;\;\; 16.4,\;\;\; 20.2,\ldots \label{jiroot}
\end{eqnarray}
again for $m=1,2$ and 3 respectively in each of the rows.

In order to calculate the energies in equation (\ref{6.3}), we need also to
know the experimental values of $R$ (the radius of the disc on which the above
quantization has been carried out) and the mass $em_{H}$ of the gauge bosons.
  If we further assume that $1/R$ is much smaller than $em_{H}$ (which is
physically reasonable being equivalent to assuming that the penetration depth
is
much smaller than the radius of the disc), then we can assume that
$|\frac{\b_{nm}}{em_{H}}|<<1$ (at least for the values of $\b_{nm}R$ in
(\ref{j0root}) and (\ref{jiroot})).  Hence the following approximation for the
variation $\delta \Omega _{nm}^{(\b )}$ of $\Omega _{nm}^{(\b )}$ under the
variation $\delta \b_{nm}$ of $\b_{nm}$ follows from $\O_{nm}^{(\b )}=
\sqrt{\b_{nm}^2 + e^{2}m^2 _{H}}$:
\begin{eqnarray}
\delta \O_{nm}^{(\b )} &=& \frac{\b_{nm} \delta \b_{nm}}
{\sqrt{\b_{nm}^2 + e^{2}m^2 _{H}}} \approx \frac{\b_{nm}}{em_{H}}\delta \b_{nm}
\label{endep}
\end{eqnarray}
If we choose a typical experimental value of $10^{4}$ for $em_{H}R$, then the
above energy dependence reduces to
\be
\delta \O_{nm}^{(\b )} \approx 10^{-4} \delta \b_{nm}, \label{exptal}
\ee
showing therefore that the energy changes are extremely minute (much smaller
than the corresponding changes in $\b_{nm}$) in this
approximation.

However, this is no longer the case if we consider a very tiny disc whose
radius
is of the order of the penetration depth itself.  Then instead of (\ref{endep})
we would have
\be
\delta \O_{nm}^{(\b )} \approx \delta \b_{nm} \label{endep2}
\ee
so that, changes in the energy are now of the same order as the changes in
$\b_{nm}$ (namely, $1/R$ times the difference of the corresponding numbers in
(\ref{j0root}) and (\ref{jiroot})).

{}From this rough analysis, we expect the effect of boundary conditions on the
spectrum to be increasingly prominent as the size of the system under
consideration becomes smaller.

\sxn{Higher Dimensions}

Edge observables and their states are consequences of gauge invariance.  They
can exist in gauge theories on manifolds with boundaries regardless of spatial
dimensionality.  The algebra of these observables is sensitive to the nature of
the gauge theory and is generally abelian for an abelian gauge theory (such as
the Maxwell theory) when
topological interactions are absent.  But it can become nonabelian even for an
abelian gauge theory in the presence
of the latter as happens with the CS term.

The abelian generalisation of the CS Lagrangian to 3+1 dimensions is the $BF$
interaction involving the two-form field $B$ and the curvature $F=dA$ of the
abelian connection $A$.  With the Maxwell term and its analogue for $B$
included in the action, the model describes the London equations and gives
a new approach to superconductivity and mass generation of vector mesons
\cite{bow,pao} differing from that based on the Higgs field.
We think therefore that there are good reasons to pay attention to this action.
In previous work \cite{pao}, we have examined it with emphasis on its edge
observables and
have shown that the latter are strikingly similar to their 2+1 analogues.  Thus
they provide a generalisation of affine Lie groups and enjoy a description
using coadjoint orbits and a topological action exactly like the WZNW action.
The latter, as shown in \cite{brai}, is the coadjoint action for the Kac-Moody
group and is known to be associated with the edge states of the CS theory
\cite{witt,csb}.

This previous work on the $BF$ system \cite{pao} had an important limitation
in that it did not investigate
possible BC's using operatorial methods.  We have seen before \cite{mcs}, and
will also see in Section 7.1, that BC's leading to edge observables in 2+1
dimensions are very special,
and require the medium outside the manifold to be vacuum.  More general BC's,
involving the substitution of other media for the vacuum, delocalise the edge
effects.

With this knowledge in mind, we reexamine the $BF$ interaction in this Section
with a solid ball ${\cal B}_{3}$ for a spatial manifold (this choice being only
for specificity).  The presence of parameters like those in (\ref{4.3}) is
established here too and their interpretation is indicated.  Next we generalise
our remarks of Section 3 and 4 regarding edge states of vortices to 't
Hooft-Polyakov monopoles \cite{mon}.

\subsxn{The BF Interaction}

The Lagrangian of interest is
\begin{eqnarray}
L & = & \int d^{3}x\, {\cal L}, \label{7.1} \\
{\cal L} & = & \frac{1}{2}\epsilon
^{\mu\nu\lambda\rho}B_{\mu\nu}F_{\lambda\rho} -\frac{1}{3\gamma}H^{\mu\nu
\lambda}H_{\mu\nu\lambda}-\frac{1}{4}F^{\mu\nu}F_{\mu\nu}, \nonumber\\
H_{\mu\nu\lambda} &=& \partial _{\mu}B_{\nu\lambda}+\partial _{\nu}B_{\lambda
\mu}+\partial _{\lambda}B_{\mu\nu}, \;\;\; \gamma >0, \nonumber\\
\epsilon ^{\mu\nu\lambda\rho}&=& \mbox{the Levi-Civita symbol with }
\epsilon ^{0123}=1.\label{7.2}
\end{eqnarray}

We first briefly review some pertinent results of \cite{pao}.

(\ref{7.1}) leads to the two first class constraints
\be
\int _{{\cal B}_{3}}d^{3}x\partial _{i}\chi ^{(0)}\Pi _{i} \approx
0,\label{7.3}
\ee
\be
-\frac{1}{2}\int _{{\cal B}_{3}}d^{3}x\partial _{j}\chi _{i}^{(1)}[P_{ji} +
2\epsilon _{ijk}A_{k}] \approx 0\label{7.4}
\ee
where $\Pi _{i}=F_{0i}+\epsilon _{ijk}B_{jk}$ and
$P_{ij}=\frac{4}{\gamma}H_{0ij}$ are fields of momenta conjugate to $A_{i}$ and
$B_{ij}$, and $\chi ^{(0)}$ and $\chi ^{(1)}$ are zero and one forms with zero
pull backs to $\partial {\cal B}_{3}$.  We can write the latter constraints as
the BC's
\begin{eqnarray}
\chi ^{(0)}|_{\partial {\cal B}_{3}}&=&\chi ^{(1)}|_{\partial {\cal B}_{3}} =
0,\nonumber\\
\chi ^{(1)}&\equiv &\chi _{i}^{(1)}dx^{i}. \label{7.5}
\end{eqnarray}
(\ref{7.5}) are found by requiring differentiability of
(\ref{7.3},\ref{7.4}).

The constraints (\ref{7.3},\ref{7.4}) generate the gauge transformations
\begin{eqnarray}
&&A  \rightarrow  A + d\chi ^{(0)},\label{7.6} \\
&&B  \rightarrow  B + d\chi ^{(1)},\label{7.7} \\
&&A\equiv A_{i}dx^{i}, \;\; B\equiv \frac{1}{2}B_{ij}dx^{i}dx^{j} \label{defn}
\end{eqnarray}
as can be shown using the usual equal time commutators
\begin{eqnarray}
{[ A_{i}(x), \Pi _{j}(y)]} & = & i\delta _{ij}\delta ^{3}(x-y)\nonumber \\
{[B_{ij}(x), P_{kl}(y)]} & = & i(\delta _{ik}\delta _{jl} -\delta _{il}\delta
_{jk})\delta ^{3}(x-y) \label{7.8}
\end{eqnarray}
these being of course the only nonvanishing commutators for these fields.  [The
wedge symbol between differential forms is being omitted in (\ref{defn}) and
below.]

The algebra of edge observables is generated by
\begin{eqnarray}
Q(\Lambda ^{(0)}) & = & \int _{{\cal B}_{3}}d^{3}x\partial _{i}\Lambda ^{(0)}
\Pi _{i} \label{7.9}\\ R(\Lambda ^{(1)}) & = & -\frac{1}{2}\int _{{\cal B}_{3}}
d^{3}x\partial _{j}\Lambda _{i}^{(1)}(P_{ji}+2\epsilon
_{ijk}A_{k}).\label{7.10}
\end{eqnarray}
They are got from (\ref{7.3},\ref{7.4}) by relaxing the boundary conditions
in (\ref{7.5}) in the usual
way.  Their nonvanishing commutators are all given by
\begin{eqnarray}
{[Q(\Lambda ^{(0)})\; ,\; R(\Lambda ^{(1)})]}& =& i\int _{{\cal B}_{3}}
d\Lambda ^{(0)}d\Lambda ^{(1)}, \nonumber\\
\Lambda ^{(1)} &\equiv & \Lambda _{i}^{(1)}dx^{i}. \label{7.11}
\end{eqnarray}

As mentioned before, operator quantisation involves BC's on $\partial {\cal
B}_{3}$ and (\ref{7.9},\ref{7.10}) are well-defined only for a particular
choice of
these BC's.  These BC's emerge when we begin diagonalising the Hamiltonian
\begin{eqnarray}
H & = & \int _{{\cal B}_{3}}d^{3}x[\frac{1}{2}(\Pi _{i} -\epsilon _{ijk}B_{jk})
^{2}+\frac{\gamma}{16}P_{ij}^{2} + \frac{1}{4}F_{ij}^{2} +\frac{1}{3\gamma}
H_{ijk}^{2}] \nonumber \\
& = & \frac{1}{2}(\Pi -2*B,\Pi -2*B)+\frac{\gamma}{8}(P,P)+\frac{1}{2}(A,*d*dA)
-\frac{2}{\gamma}(B,*d*dB),\nonumber
\end{eqnarray}
\be
\label{7.12}
A:=A_{i}dx^{i},\; B:=\frac{1}{2}B_{ij}dx^{i}dx^{j},\; \Pi :=\Pi _{i}dx^{i},\;
P:=\frac{1}{2}P_{ij}dx^{i}dx^{j}.
\ee
[As before $d$ and $*$ refer respectively to the exterior derivative and the
Hodge star on the spatial manifold (see (\ref{hodge*}) to (\ref{defd}) for the
relevant definitions).]  Here we have introduced a scalar
product on the space of $p$-forms ($p=1,2$):
\be
(\alpha ^{(p)},\beta ^{(p)}):= \int _{{\cal B}_{3}}\bar{\alpha}^{(p)}*\beta
^{(p)}. \label{scprod}
\ee

Equation (\ref{7.12}) shows that diagonalisation of $H$ involves that of
$*d*d$ in the space of one- and two-forms.  So we
must first study the domains of self-adjointness of this operator on the space
of one-forms as well as of two-forms.

These domains can be found using standard rules characterising a domain, and
also locality of the boundary conditions, the analysis being much the same as
in
\cite{mcs,higgs}.  The possible domains ${\cal D}_{\lambda ,\mu}$ are
characterised by two parameters $\lambda$ and $\mu$ and are given by
\be
{\cal D}_{\lambda ,\mu} = \{ A, B|(*^{(3)}dA)|_{\partial {\cal B}_{3}} =
\lambda
*^{(2)}(A|_{\partial {\cal B}_{3}});(*^{(3)}dB)|_{\partial {\cal B}_{3}} = -
\mu *^{(2)}(B|_{\partial {\cal B}_{3}}
);\lambda , \mu \in \mbox{l$\!\!$R},\; \lambda ,\mu\geq 0 \},\label{7.13}
\ee
where the superscripts $^{(3)}$ and $^{(2)}$ on the $*$'s indicate the fact
that
these are defined respectively on the three-dimensional ball and its
two-dimensional boundary.  We shall omit these superscripts whenever there
is no source for confusion about the dimension on which the $*$ is defined.
The requirements $\lambda ,\mu \geq 0$ arise from the requirement of
non-negativity of the Hamiltonian (\ref{7.12}).  The quickest way to see how
this comes about is by writing the terms in $H$ which are not manifestly
positive-semidefinite as follows:
\be
 \frac{1}{2}(A,*d*dA) -\frac{2}{\gamma}(B,*d*dB)=\frac{1}{2}(dA,dA)
+\frac{2}{\gamma}(dB,dB) -\int _{\partial {\cal B}}[\frac{1}{2}A(*^{(3)}dA)
+\frac{2}{\gamma}B(*^{(3)}dB)]. \label{nonneg}
\ee
On using (\ref{7.13}) in the RHS of above, we now get $\lambda ,\mu \geq 0$ in
order that this expression be non-negative.

The physical meaning of $\lambda$ can be deduced as before by
considering an arrangement with ${\cal B}_{3}$ surrounded by a superconductor,
in general different from what is inside ${\cal B}_{3}$.  We can describe the
latter using the Landau-Ginzburg or the $BF$ formalism \cite{bow,pao}, the
latter being more
appropriate for this Section, the system in ${\cal B}_{3}$ having been
described
in that way.  So let $A$ be the electromagnetic potential as before and $B$
the two-form potential for the current $J$ outside the ball:
\be
J^{\mu} =-\epsilon ^{\mu\nu\lambda\rho}\partial
_{\nu}B_{\lambda\rho}.\label{curr}
\ee
 The constants in the Lagrangian and
field equations can now be different from (\ref{7.1},\ref{7.2}) if the medium
outside is
different.  We will therefore distinguish them by an additional prime.  As
in the derivation of (\ref{5.1},\ref{5.2}), here too we assume
a static distribution of fields in the exterior so that the time derivatives of
electric/magnetic fields and of the currents are assumed to be negligible.  We
then get the following equations from the
variation of the Lagrangian with respect to $A$ and $B$:
\begin{eqnarray}
&&-\frac{1}{3}\epsilon ^{\mu\nu\lambda\rho}H_{\lambda\mu\nu}+ \partial
_{\sigma}F^{\sigma\rho}=0,\label{onnu}\\
&&\epsilon ^{\mu\nu\lambda\rho}F_{\lambda\rho}+\frac{4}{\gamma '}\partial
_{\rho}H^{\rho\mu\nu}=0.\label{rendu}
\end{eqnarray}
On using also the assumption that
the electric and magnetic fields and the current are static, (\ref{onnu}) and
(\ref{rendu}) give the following equations (besides others):
\begin{eqnarray}
d*dA +\frac{\gamma '}{2}P & = & 0\;\; ,\;\; x\in
\mbox{l$\!\!\!$R$^{3}$}\backslash
{\cal B}_{3} ,\label{7.14}\\
*d(A -\frac{1}{2}*P) & = & 0\;\; ,\;\; x\in \mbox{l$\!\!\!$R$^{3}$}\backslash
{\cal B}
_{3} .\label{7.15}
\end{eqnarray}
Here we have used our definitions $P_{ij}=\frac{4}{\gamma '}H_{0ij},\;
P=\frac{1}{2}P_{ij}dx^{i}dx^{j}$ and the
fact that $J^{i}$  is the same as $\epsilon ^{ijk}H_{0jk}$ in view of
(\ref{curr}).

{}From (\ref{7.15}) it follows that $A -\frac{1}{2}*P =d\Phi$.  Thus
(\ref{7.14}) implies the equations
\begin{eqnarray}
d*dA +\gamma '*(A-d\Phi ) & = & 0 ,\label{7.16}\\
d*(A -d\Phi ) & = & 0 ,\label{7.17}
\end{eqnarray}
(\ref{7.17}) following from (\ref{7.16}) by applying $d$.

Exactly as in the arguments following (\ref{5.1},\ref{5.2}), it can be shown
that the above two equations imply that $A-d\Phi$ decays like $\exp
[-\sqrt{\gamma '}(r-R)]$ outside ${\cal B}_{3}$.  Equation (\ref{7.16}) being
an equation
for two-forms, we now integrate it over a
two-surface $S$ to get the BC's.  The latter is chosen as follows.  Let ${\cal
C}_{1}$ be a small segment on $\partial {\cal B}_{3}$.  Then $S$ has the shape
of a thin strip with ${\cal C}_{1}$ and another segment ${\cal C}_{2}$ as
boundaries (see figure 4).

\begin{figure}[hbt]
\begin{center}\mbox{\psannotate{\psbox{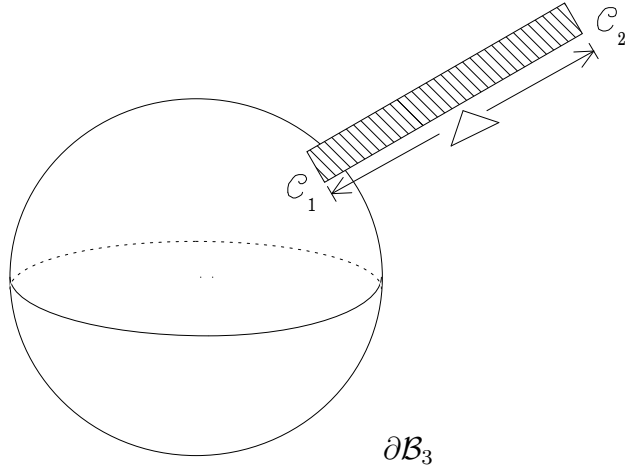}}
{\at(5\pscm;0\pscm){$\partial {\cal B}_{3}$}}}
\end{center}
\caption{Figure showing a strip coming out from the boundary of ball}
\end{figure}

On integration, we find,
\be
\int _{{\cal C}_{2}}*dA - \int _{{\cal C}_{1}}*dA +\gamma '\int
_{S}*(A-d\Phi) = 0 .\label{7.16a}
\ee

Let $r,\theta , \varphi$ be polar coordinates centered at the middle of ${\cal
B}_{3},\; \partial {\cal B}_{3}$ having the value $R$  and ${\cal C}_{2}$ the
value $R+\Delta$ for $r$.  So ${\cal C}_{2}$ is at a distance $\Delta$ from
$\partial {\cal B}_{3}$.  Choosing $\Delta$ so that $\sqrt{\gamma '}
\Delta >>1$, we thus find,
\be
\int _{{\cal C}_{1}}*dA = \gamma '\int _{S}*(A-d\Phi ). \label{7.17a}
\ee
Now, ${\cal C}_{1}$ being small, a further approximation is possible.  If
$B_{i}$ is the tangential component of the magnetic field $\vec{B}$ along
${\cal C}_{1}$ and $B_{i}(P)$ its value at
a point $P$ of ${\cal C}_{1}$, (\ref{7.17a}) approximately gives
\be
B_{i}(P) = -\sqrt{\gamma '}\epsilon _{ij}(A_{j}-\partial _{j}\Phi)(P)
\label{7.18}
\ee
Here $i$ labels the direction of this tangent while $j$ labels the
direction perpendicular to this tangent on the surface $\partial {\cal B}_{3}$.
This equation can be written as
\be
\label{com}
(*^{(3)}dA)|_{\partial {\cal B}_{3}}=\sqrt{\gamma '}*^{(2)}[(A-d\Phi
)|_{\partial {\cal B}_{3}}]
\ee
where the $*^{(j)}$'s here have the same meaning as in (\ref{7.13}).  The first
minus sign in the right hand side of (\ref{7.18}) is absent in (\ref{com})
because $\partial {\cal B}_{3}$ has opposite orientations when viewed as
boundary of domains outside and inside ${\cal B}_{3}$.  Equation (\ref{com}) is
 the same as the boundary condition on $A$ in (\ref{7.13}),
$\lambda$ having the interpretation
\be
\lambda = \sqrt{\gamma '}\label{7.19}
\ee
provided also that $d\Phi |_{\partial {\cal B}_{3}}=0$.

In other words, $\frac{1}{\lambda}$ is the penetration depth.

It merits emphasis that the sign of $\lambda$ required for this interpretation
is the same as the sign demanded by non-negativity of energy. [See
(\ref{nonneg}) and what follows it.]

There remains the condition on the two-form $B$ in (\ref{7.13}).  It is to be
regarded as a boundary condition for solving the equations
(\ref{7.14},\ref{7.15}) and the other equations that arise from
(\ref{onnu},\ref{rendu}).

Gauss laws generate gauge transformations with test functions vanishing at
$\partial {\cal B}_{3}$.  The BC's (\ref{7.13}) are compatible with these
transformations.  They are not however compatible with the transformations due
to the charges (\ref{7.9},\ref{7.10}) with test functions nonvanishing at
$\partial {\cal
B}_{3}$ unless $\lambda =\mu = 0$.  For this reason, it is only for this
special
choice of BC's that the edge observables exist as operators with a well-defined
action on the Fock states.  For other choices of BC's, they must be thought of
as spontaneously broken.

The detailed quantisation of (\ref{7.1}) using the BC's (\ref{7.13}) is
possible
for $\lambda =\mu =0$, although we have not done so.  [See reference \cite{pao}
in this regard.]  For $\lambda , \mu \neq
0$, however, quantisation meets with the same difficulties as in the
Maxwell-Chern-Simons case.

We also finally remark that a similar analysis in 3+1 dimensions can be
performed using the
Landau-Ginzburg, that is the gauged Higgs field, description of
superconductors.

\subsxn{Edge States for Monopoles}

The monopoles under consideration here are the 't Hooft-Polyakov monopoles in
the Georgi-Glashow model \cite{mon}.  In the $A_{0}=0$ gauge, they are
constituted of an
$SO(3)$ connection $A=A_{i}^{\alpha}dx^{i}\frac{1}{2}\tau _{\alpha}\equiv A_{i}
dx^{i}$ and a Higgs field $\Phi = \varphi _{\alpha}\frac{1}{2}\tau _{\alpha}$
transforming by the
$SO(3)$ triplet representation, $\tau _{\alpha}$ being Pauli matrices.  Let us
think of the monopole as approximately confined to a ball ${\cal B}_{3}$ of
radius $R$ so that
\be
D_{i}\Phi = \partial _{i}\Phi + ie[A_{i}, \Phi ] =0\mbox{ for } r >
R, \label{7.20}
\ee
the polar coordinates $r,\theta,\varphi$ having origin at the center of ${\cal
B}_{3}$ as before.  The condition (\ref{7.20}), as is well-known \cite{mon}, is
a consequence of requiring
that the monopole energy is finite.  For the static monopole, $A$ and $\Phi$
become $A^{(0)} = A^{(0)3}\frac{1}{2}\tau _{3}$ and  $\Phi ^{(0)}= \Phi ^{(0)3}
\frac{1}{2}\tau _{3}$
in the
U-gauge, $A^{(0)3}$ being the potential for the static Dirac monopole field and
$\Phi ^{(0)3}$ a nonzero constant.

Consider the field fluctuations $a_{\mu},\; \varphi ,\; \varphi ^{*}$ around
the static solution defined by
\begin{eqnarray}
A_{\mu} & = & [A_{\mu}^{(0)3} + a_{\mu}]\frac{1}{2}\tau _{3}, \nonumber \\
\Phi & = & \Phi ^{(0)3}\frac{1}{2}\tau _{3} + \varphi \frac{1}{2}\tau _{+} +
\varphi ^{*}\frac{1}{2}\tau _{-} \nonumber\\
&\equiv & \Phi ^{(0)}+\delta \Phi ,\nonumber\\
\tau _{\pm} &=&\tau _{1}\pm i\tau_{2}.\label{7.21}
\end{eqnarray}
They are compatible with (\ref{7.20}) if
\be
\partial _{i}\phi +ieA_{i}\phi \equiv D_{i}\phi =0 \mbox{ for } r>R.
\label{7.215}
\ee

Now consider the dynamics of a complex field $\varphi$ with $U(1)$ charge $e$
in
interaction with a $U(1)$ connection in the region $\check{\cal B}_{3}$ outside
the monopole.  These excitations
are considered with their energy localised away from the monopole.

The dynamics of $a_{\mu}$ and $\varphi$ in $\check{\cal B}_{3}$ can be
approximated by the usual Lagrangian $L$ of a charged scalar field:
\begin{eqnarray}
L & = & \int _{\check{\cal B}_{3}}d^{3}x\, {\cal L} ,\nonumber \\
{\cal L} & = & -(D_{\mu}\varphi)^{*}(D^{\mu}\varphi) -\frac{1}{4}(\partial
_{\mu}a_{\nu}-\partial _{\nu}a_{\mu})(\partial ^{\mu}a^{\nu}-\partial
^{\nu}a^{\mu}),\nonumber\\
D_{\mu}\varphi &\equiv & \partial _{\mu}\varphi +ie(A_{\mu}^{(0)3}+a_{\mu})
\varphi .\label{7.23}
\end{eqnarray}
It has the Gauss law constraint
\begin{eqnarray}
{\cal G}(\Lambda ^{(0)}) |\cdot \rangle & = & 0 ,\nonumber \\
{\cal G}(\Lambda ^{(0)}) & = & \int _{\check{\cal B}_{3}}d^{3}x\Lambda ^{(0)}
\partial _{i}\Pi _{i} +\int _{\check{\cal B}_{3}}d^{3}x
\Lambda ^{(0)}ie(\varphi \pi _{\varphi} -\varphi ^{*}\pi
_{\varphi}^{*}),\nonumber\\
 \Lambda ^{(0)}|_{\partial {\cal B}_{3}}&=0 =& \Lambda
^{(0)}|_{\infty}, \label{7.24}
\end{eqnarray}
$|\cdot \rangle$ being any physical state and $\Pi_{i}$ and $\pi_{\varphi}$
being the fields conjugate to $a_{i}$ and $\varphi$.  The field $\Pi_{i}$ is of
course the electric field.

The edge observables localised at the boundary $\partial {\cal B}_{3}$ of the
monopole are obtained from (\ref{7.24}) by partial integration:
\be
Q(\Lambda ) = -\int _{\check{\cal B}_{3}}d^{3}x\partial _{i}\Lambda \Pi _{i} +
\int _{\check{\cal B}_{3}}d^{3}x\Lambda ie(\varphi \pi _{\varphi} -\varphi ^{*}
\pi _{\varphi}^{*}). \label{7.25}
\ee
Here $\Lambda |_{\partial {\cal B}_{3}}$ need not vanish while $\Lambda
|_{\infty}$ is still zero.  [Nonzero $\Lambda |_{\infty}$ is associated with
the conventional charge operator.]

The algebra of these observables is abelian:
\be
{[Q(\Lambda ), Q(\bar{\Lambda})]} = 0. \label{7.26}
\ee
These observables can be interpreted as the infinitely many conserved charges
of $L$ localized at $\partial {{\cal B}_{3}}$.

The system in $\check{\cal B}_{3}$ is not in a broken phase as regards the
$U(1)$ corresponding to $\tau _{3}$.  We cannot therefore write $\varphi$ =
constant ($\neq 0)\times e^{i\Psi}$ in $\check{\cal B}_{3}$ and thereby get an
approximate quadratic Lagrangian for this system.  A simple gauge invariant
approach for the investigation of possible BC's compatible with the Gauss law
is not therefore available.  This means that we are unable to generalise our
previous work on BC's and display the above conditions on $\Lambda |_{\partial
{\cal B}_{3}}$ as being associated with a special choice of these BC's.

\centerline{ {\bf Acknowledgements}}

\nopagebreak
We thank Giuseppe Bimonte, T.R.Govindarajan, Dimitra Karabali, Ajit Mohan
Srivastava and Paulo Teotonio-Sobrinho for several discussions.
The work of A.P.B., L.C. and E.E.
was supported by the Department of Energy, U.S.A., under contract number
DE-FG02-85ER40231.


\begin{thebibliography}{99}

\bibitem{qhe}  B.T. Halperin, Phys. Rev. \underline{B25} (1982) 2185; G.
Morandi, Quantum Hall Effect, Monographs and Textbooks in Physical
Science, Lecture Notes Number 10, Bibliopolis, Naples (1988); M. Stone,
Ann. Phys. \underline{207} (1991) 38; X.G. Wen,
Int. Jour. Mod. Phys. \underline{B6} (1992) 1711; A.P. Balachandran and A.M.
Srivastava, Syracuse University and Univesity of Minnesota preprint SU-4228-492
TPI-MINN-91-38-T (1991) and hepth/9111006.
\bibitem{witt} E.Witten, Comm. Math. Phys. \underline{121} (1989) 351;
S.Elitzur, G.Moore, A.Schwimmer and N.Seiberg, Nucl. Phys. \underline{B326}
(1989) 108 and references therein.
\bibitem{csb} A.P.Balachandran, G.Bimonte, K.S.Gupta and A.Stern, Int.\@ J.\@
Mod.\@ Phys.\@ \underline{A7} (1992) 4655, 5855 and references therein.
\bibitem{mcs} A.P.Balachandran, L.Chandar, E.Ercolessi, T.R.Govindarajan and
R.Shankar, Int. J. Mod. Phys. \underline{A9} (1994) 3417.
\bibitem{higgs} L.Chandar and E.Ercolessi, Syracuse University preprint
SU-4240-575 (1994) and hep-th/9408047.
\bibitem{pao} A.P.Balachandran and P.Teotonio-Sobrinho, Int.\@ J.\@ Mod.\@
Phys.\@ \underline{A8} (1993) 723; \underline{A9} (1994) 1569;
A.P.Balachandran, G.Bimonte and P.Teotonio-Sobrinho, Mod.\@ Phys.\@ Lett.\@
\underline{A8} (1993) 1305.
\bibitem{dyo} E.Witten, Phys.\@ Lett.\@ \underline{86B} (1979) 283.
\bibitem{jack} R.Jackiw and C.Rebbi, Phys.\@ Rev.\@ \underline{D13} (1976)
3398; R.Jackiw and J.R.Schrieffer, Nucl. Phys. \underline{B190} (1981)
253.
\bibitem{camp}  A.P.Balachandran,
``Gauge Symmetries, Topology and Quantization'', AIP conference Proceedings
317, Fifth Mexican School of Particles and Fields, Guanajuato, Mexico, 1992,
Editors J.L.Lucio and M.Vargas (1994) pp. 1-81.
\bibitem{call} C.G.Callan, R.F.Dashen and D.J.Gross, Phys. Lett.
\underline{63B} (1976) 334; R.Jackiw and C.Rebbi, Phys. Rev. Lett.
\underline{37} (1976) 172.
\bibitem{book} Cf. A.P.Balachandran, G.Marmo, B.S.Skagerstam and A.Stern,
``Classical Topology and Quantum States'', World Scientific (1991) and
references therein.
\bibitem{schw} For a review, see for example N.S.Manton, Ann.\@ Phys.\@ (N.Y.)
\underline{159} (1985) 220.
\bibitem{vo} For a review, see for example P.Goddard and D.Olive, Int.\@ J.\@
Mod.\@ Phys.\@ \underline{A1} (1986) 303.
\bibitem{cs} J.F.Schonfeld, Nucl.\@ Phys.\@ \underline{B185} (1981) 157;
S.Deser, R.Jackiw and S.Templeton, Phys.\@ Rev.\@ Let.\@ \underline{48} (1982)
975; Ann.\@ Phys.\@ \underline{140} (1982) 372.
\bibitem{unp} A.P.Balachandran, G.Bimonte, K.S.Gupta and A.Stern (unpublished).
\bibitem{bow} M.J.Bowick, S.B.Giddings, J.A.Harvey, G.T.Horowitz and
A.Strominger, Phys. Rev. Lett. \underline{61} (1988) 2823; J.A.Minahan and
R.C.Warner, Florida Preprint UFIFT-HEP-89-15 (1989); T.J.Allen, M.Bowick
and A.Lahiri, Mod. Phys. Lett. \underline{A6} (1991) 559;  A.Lahiri,
Mod. Phys. Lett. \underline{A8} (1993) 2403; and references therein.
\bibitem{brai} B.Rai and V.G.J.Rodgers, Nucl.\@ Phys.\@ \underline{B341} (1990)
119; A.Yu.\@ Alekseev and S.L.Shatashvili, Mod.\@ Phys.\@ Lett.\@
\underline{A3} (1988) 1551; Nucl.\@ Phys.\@ \underline{B323} (1989) 719.
\bibitem{mon} For a review, see for example P.Goddard and D.Olive, Rep.\@ Prog.
\@ Phys.\@ \underline{41} (1978) 1357.

\end{thebibliography}
\end{document}